\documentclass[11pt]{article}

\usepackage{geometry}
\usepackage[printwatermark]{xwatermark}
\usepackage{xcolor}
\usepackage{authblk}

\usepackage{lineno}
\usepackage{parskip}

\usepackage{setspace}
\usepackage{amsmath,amssymb}
\usepackage{gensymb}
\usepackage{graphicx}

\title{Multi-messenger Astrophysics: Harnessing the Data Revolution}
\date{\today}
\author[1]{Gabrielle Allen}
\affil[1]{National Center for Supercomputing Applications, University of Illinois at Urbana-Champaign}
\author[2]{Warren Anderson}
\affil[2]{Center for Gravitation, Cosmology and Astrophysics, University of Wisconsin-Milwaukee}
\author[3]{Erik Blaufuss}
\affil[3]{Department of Physics, University of Maryland}
\author[4]{Joshua S. Bloom} 
\affil[4]{Department of Astronomy, University of California, Berkeley}
\author[2]{Patrick Brady}
\author[5,6]{Sarah Burke-Spolaor}
\affil[5]{Department of Physics and Astronomy, West Virginia University}
\affil[6]{Center for Gravitational Waves and Astronomy, West Virginia University}
\author[7,8]{S. Bradley Cenko}
\affil[7]{NASA Goddard Space Flight Center}
\affil[8]{Joint Space-Science Institute}
\author[9]{Andrew Connolly} 
\affil[9]{Department of Astronomy, University of Washington}
\author[10]{Peter Couvares}
\affil[10]{LIGO Laboratory, California Institute of Technology}
\author[11]{Derek Fox}
\affil[11]{Department of Astronomy and Astrophysics, Pennsylvania State University}
\author[12]{Avishay Gal-Yam}
\affil[12]{Department of Particle Physics and Astrophysics, Weizmann Institute of Science}
\author[13,8]{Suvi Gezari}
\affil[13]{Department of Astronomy, University of Maryland}
\author[14]{Alyssa Goodman} 
\affil[14]{Harvard-Smithsonian Center for Astrophysics}
\author[15]{Darren Grant}
\affil[15]{Department of Physics, University of Alberta}
\author[16]{Paul Groot}
\affil[16]{Department of Astrophysics, Radboud University}
\author[17]{James Guillochon}
\affil[17]{Institute for Theory and Computation, Harvard-Smithsonian Center for Astrophysics}
\author[18]{Chad Hanna}
\affil[18]{Department of Physics, Pennsylvania State University}
\author[19,20]{David W. Hogg} 
\affil[19]{Center for Cosmology and Particle Physics, New York University}
\affil[20]{Center for Computational Astrophysics, Flatiron Institute}
\author[21]{Kelly Holley-Bockelmann}
\affil[21]{Department of Physics and Astronomy, Vanderbilt University}
\author[22,23]{D. Andrew Howell} 
\affil[22]{Las Cumbres Observatory}
\affil[23]{Department of Physics, University of California, Santa Barbara}
\author[2]{David Kaplan} 
\author[24]{Erik Katsavounidis}
\affil[24]{LIGO Laboratory, Massachusetts Institute of Technology}
\author[25,26]{Marek Kowalski}
\affil[25]{Deutsches Elektronen-Synchrotron (DESY)}
\affil[26]{Institute of Physics, Humboldt University of Berlin}
\author[27]{Luis Lehner}
\affil[27]{Perimeter Institute for Theoretical Physics}
\author[28]{Daniel Muthukrishna}
\affil[28]{Institute of Astronomy, University of Cambridge}
\author[29]{Gautham Narayan} 
\affil[29]{Space Telescope Science Institute}
\author[29,30]{J.E.G. Peek}
\affil[30]{Johns Hopkins University}
\author[29]{Abhijit Saha}
\author[3,8]{Peter Shawhan}
\author[31]{Ignacio Taboada} 
\affil[31]{School of Physics, Georgia Institute of Technology}

\begin{document}

\maketitle

\noindent\rule{\linewidth}{\arrayrulewidth}

\begin{abstract}
The past year has witnessed discovery of the first identified
  counterparts to a gravitational wave transient (GW\,170817A) and a
  very high-energy neutrino (IceCube-170922A). These source
  identifications, and ensuing detailed studies, have realized
  longstanding dreams of astronomers and physicists to routinely carry
  out observations of cosmic sources by other than electromagnetic
  means, and inaugurated the era of ``multi-messenger'' astronomy.
  While this new era promises extraordinary physical insights into the
  universe, it brings with it new challenges, including: highly
  heterogeneous, high-volume, high-velocity datasets; globe-spanning
  cross-disciplinary teams of researchers, regularly brought together
  into transient collaborations; an extraordinary breadth and depth of
  domain-specific knowledge and computing resources required to
  anticipate, model, and interpret observations; and the routine need
  for adaptive, distributed, rapid-response observing campaigns to
  fully exploit the scientific potential of each source.
  We argue, therefore, that the time is ripe for the community to
  conceive and propose an Institute for Multi-Messenger Astrophysics
  that would coordinate its resources in a sustained and
  strategic fashion to efficiently address these challenges, while
  simultaneously serving as a center for education and key supporting
  activities. 
  In this fashion, we can prepare now to realize the bright future
  that we see, beyond, through these newly opened windows onto the
  universe.\\
  
  {\it This document is a summary from an NSF-supported workshop held on 23-24 May 2018 at the University of Maryland. Attendees were charged to consider how multi-messenger astrophysics can be enabled by enhanced cyberinfrastructure. This workshop summary outlines some of the opportunities, significant challenges, and community needs of MMA over the next decade. }

\end{abstract}

\noindent\rule{\linewidth}{\arrayrulewidth}
\clearpage

\tableofcontents
\clearpage

\section{Executive Summary}
\label{s:exec_summary}
Within the past year, the discovery of the first electromagnetic counterparts to sources of gravitational waves and high-energy neutrinos, achieved following decades of National Science Foundation and NASA support, have realized astronomy's longstanding dream of observing cosmic sources by means other than light. This new dawn of multi-messenger astrophysics, the first such discoveries since the historic neutrino detections of the Sun and Supernova\,1987A decades ago, offers unique opportunities for insight into the physical universe, while at the same time posing distinct challenges. We think it is imperative to call on the community to confront these challenges, and seize these opportunities, with a sustained and strategic investment in applied theory, data science algorithms and inference methods, community building, and cyberinfrastructure for multi-messenger astrophysics (MMA). 

Returns on these investments are likely to include: Solving the mystery of the highest-energy cosmic rays and identifying their cosmic particle accelerators; identifying the formation and growth mechanisms for black holes across the mass spectrum and throughout the universe; characterizing the properties of hot and cold baryonic matter at super-nuclear densities; and testing our physical theories under extreme conditions not reproducible in any terrestrial laboratory. 

\textbf{The Multi-messenger Opportunity:}
Discovery of the gamma-ray burst and kilonova / afterglow counterparts to the binary neutron star gravitational-wave event GW170817 (Abbott et al. 2017, ApJL, 848, L12), and identification and characterization of the blazar TXS 0506+056 associated with the IceCube-170922A high-energy neutrino (The IceCube Collaboration et al. 2018, Science, 361, 1378), have brought us to a unique moment. Two fundamentally new windows onto the Universe have been opened, and continued operation and expansion of ultra-high-energy cosmic ray facilities promise more to come. At the same time, next-generation survey facilities across the electromagnetic spectrum, from radio to optical to gamma-rays, will supply a flood of data with volumes and velocities orders of magnitude greater than astronomy has previously known.  

Full exploitation of the MMA opportunity thus requires a new convergence of astrophysical theories and observation, computer science, statistics, and data science. The multitude of cross-disciplinary global teams calls for easily-deployed, flexible platforms and tools for efficient collaboration and communication. Rapid and responsive communications between teams could increase the speed, effectiveness, and efficiency of observations, computations, and source modeling. The coordination of efforts to develop and improve tools, test and deploy analysis techniques, and manage the interface between observations and theory would benefit the greater community and improve scientific output by reducing duplication and promoting collaboration.
 
In particular, it is clear that significant computing resources will be required to rapidly analyze and interpret multi-messenger data sets that are heterogeneous, distributed, and sometimes proprietary. Data science innovations will be needed, including robust scaling of algorithms to process large, high-throughput data sets in real-time; to predict and prepare for moments of high data churn; and to carry out spectral time-series learning from multidimensional, heterogeneous, noisy, and sparse data. At the same time, astrophysicists are modestly trained in data science, and often unfamiliar with the latest advances in computational and statistical methods in this era where such abilities are so critical in astronomy and beyond.

\textbf{An Institute for Multi-messenger Astrophysics:}
A sustained and strategic effort will be required to bridge these gaps in expertise and infrastructure. 
Such effort must bring together researchers from all fronts
to further develop promising strategies, conceive of new ones to face the continued challenges of more refined data from multiple sources, and exploit
them to achieve the truly transformative science that MMA is uniquely positioned to drive. In addition to theoretical ideas, models, and novel data analysis
which can combine information for the multitude of sources and types, the community should  develop and promulgate effective software and communications solutions, and should deploy solutions in a timely and effective fashion so as to fully realize the promise of this new era.

%to develop and promulgate effective software and communications solutions, and to deploy solutions in a timely and effective fashion so as to realize the promise of this new era.
%
% PRB: making the case for the Center
%
\emph{The workshop attendees judged that an Institute for Multi-messenger Astrophysics would uniquely enable the effort required to address the long-term challenges in this emerging field.}
% CH and PRB
No individual investigator or observing facility can address these needs alone. 
Such an institute (possibly geographically distributed) is essential to provide a community-oriented cyberinfrastructure platform for multi-messenger astrophysics.
We envision an institute that will support the teams performing the science by facilitating efficient, multi-disciplinary collaborations across the multiple messengers and with computer and data scientists.
The institute will supplement, not supplant, existing activities to enable new, ground-breaking science.
It will draw strength from the richness of the scientific data from individual observing facilities supported by the National Science Foundation, NASA, and other agencies worldwide, and enable those data to be intelligently combined for even greater scientific impact. 
% PRB: end making case for a Center
It will allow for efficient transfer of expertise, and dissemination of tools and techniques throughout the community. 
It will also provide training opportunities in modern data science, computing, and software development practices for young scientists.

Through an institute, we can realize the promise of this new era of multi-messenger astrophysics---to not simply open new windows onto the cosmos, but to fully harness the data they yield, and extract the insights they provide into the greater universe beyond.

\section{Introduction}
\label{s:intro}
Multi-messenger astrophysics (MMA) is one of the most exciting areas of scientific discovery at the present time, with a very promising future. MMA naturally spans two of National Science Foundation's ``Big Ideas'': \emph{Windows on the Universe} and \emph{Harnessing the Data Revolution}.  By combining information from fundamentally different messengers (electromagnetic waves/photons, high-energy charged particles, neutrinos, and gravitational waves), we can obtain complementary views of energetic astronomical events and objects that, in turn, provide deep insights into astrophysics and enable tests of fundamental physics. Each messenger carries information from the astronomical event through a different physical process and combining all messengers can therefore provide the best understanding of the source. 

At the same time, MMA presents particular challenges.  It requires close coordination among disparate groups using some of the most complex scientific instruments ever built. 
Each new generation of observatories brings a substantially increased data volume and number of sources detected, which will in turn require more judicious observing strategies and more sophisticated statistical analysis and modeling to make optimal use of the them.

Participants at the May 2018 workshop included scientists representing a cross-section of astronomical facilities, computational projects, computer science and cyberinfrastructure expertise to examine the challenges and opportunities in multi-messenger astrophysics. We considered all messengers and all time scales, from simultaneous coordinated observing to carefully scheduled follow-up observing to the use of archival data sets, as well as the required characterization and necessary modeling for maximum scientific payoff.  We looked at scientific opportunities and challenges both in the near term (next few years) and over the next $\sim$20 years. In this workshop summary, we describe the outstanding opportunities in this rapidly-changing field and the need to build support and consensus around important objectives, including collaboration, coordination, cyberinfrastructure, cybersecurity, ``big data'' analysis, diversity, and sustainability.

\section{Opportunities for Multi-Messenger Astrophysics}
\label{s:opportunities}
\subsection{Capabilities of Modern Observing Facilities}

The last year has seen a new dawn of multi-messenger astrophysics, with the first electromagnetic counterparts to sources of gravitational waves (GW170817 / GRB 170817A / SSS17a / AT 2017gfo) and high-energy neutrinos (IceCube 170922A and TXS 0506+056), breaking a drought that has held since the discovery of MeV neutrinos from the Sun and SN1987A first established the power of a multi-messenger approach to understanding the universe.
Over the next 1-2 decades we can anticipate commissioning and continued operations of facilities observing the universe via a range of non-electromagnetic means: 
\begin{itemize}
\item Ground-based gravitational wave (GW) interferometers will grow to a five-detector network, operating at or beyond second-generation design sensitivity and capable of localizing transient gravitational-wave sources (at Hz-–kHz frequencies) to few-deg$^{2}$ regions of sky in real time.
\item Pulsar timing arrays (PTA) will operate as a Galaxy-scale gravitational wave detector in the nHz--$\mu$Hz frequency range.
\item MeV neutrino facilities will monitor for neutrino detection of core-collapse supernovae out to the Andromeda galaxy.
\item TeV-–PeV neutrino detectors will be capable of identifying Galactic and extra-galactic  sources that contribute to the observed IceCube diffuse neutrino flux.
\item EeV-scale neutrino detection experiments will seek to detect the cosmogenic neutrinos created by interactions of $\sim$EeV cosmic rays with the cosmic microwave background.
\item Cosmic ray facilities will seek to identify the sources of ultra-high-energy (UHE) cosmic rays directly, and to characterize the composition and energy spectrum of these cosmic rays. 
\item Space-based gravitational wave interferometers (e.g. LISA) will probe the $10^{-4}-10^{-2}$ Hz regime of gravitational waves, sensitive to the cosmic merger history of SMBHs, continuous emission from Galactic close white dwarf (WD) binaries, and the inspiral of compact objects onto SMBHs (known as EMRIs).

%Did we want to include dark matter detection (either direct or indirect) here?
\end{itemize}

Moreover, the decade will be rich with preparatory studies and theoretical work in anticipation of next-generation facilities.  Simultaneously, multi-wavelength survey facilities will be generating vast quantities of correlative data across the electromagnetic (EM) spectrum. 
In consequence, the decade promises to be a golden era for multi-messenger astrophysics, \emph{if} the data from these diverse observatories can be optimally obtained, combined and interpreted.

\subsection{Existing Cyberinfrastructure}
\label{ss:existing_cyberi}

Current MMA infrastructure has achieved notable successes, but will not suffice to deliver all of the possible science. Existing infrastructure with applications to MMA science were discussed in the workshop, including the Supernova Early Warning System (SNEWS), Astrophysical Multimessenger Observatory Network (AMON), Gamma-Ray Coordinates Network (GCN), LIGO-Virgo Collaboration (LVC) Alerts, Transient Name Server (TNS), Supernova Exchange (SNEx), the Astronomers' Telegram (ATEL), Arizona-NOAO Transient Alert and Response to Events System (ANTARES), VOEvent, and Open Astronomy Catalogs.
The infrastructure for static catalogs of the sky, including NASA/NED and Vizier/SIMBAD, is well established and robust. As presently supported they remain capable of rapidly ingesting new catalogs (e.g., Gaia DR2) and enabling redistribution and cross-correlation.  Continuation of these capabilities is critical for MMA science and source identification.

\subsection{Science That Can Be Enabled by Enhanced Cyberinfrastructure}
\label{ss:enhancement}

Realizing the full potential for MMA over the coming decade and beyond  will require prompt and intensive efforts by a broad-ranging group of physicists, astronomers, astrophysics facilities, data science specialists, and software/infrastructure engineers. These efforts will need to address the infrastructure required in order to collect and process time-critical data from observing facilities, but also to synthesize and jointly interpret such multi-messenger data.

The potential payoffs from investments in advanced cyberinfrastructure and robust coordination are great.  Example scenarios discussed in the workshop include:
\begin{itemize}
\item In the case of a relatively close binary neutron star inspiral, rapid data analysis can identify the event \emph{before} the neutron stars merge and provide a rough location in the sky.  Coordinated observations by a heterogeneous set of telescopes in orbit and on the Earth cover the sky region at the merger time to locate the counterpart, then home in to characterize the event from its prompt and afterglow emissions, including a characteristic ``kilonova'', over a wide range of wavelengths.  Detailed comparison with directed modeling reveals the physical properties of the progenitor system and tests the nuclear equation of state.
\item An energetic event, not initially seen by EM observatories, produces high-energy neutrinos which are detected and point back to a particular galaxy cluster.  Triggered deep optical and X-ray observations reveal the exact location of the source galaxy, supported by archival optical data covering the galaxy's nuclear region.  Full multi-wavelength characterization identifies the transient as a tidal disruption event---the shredding of a star by a black hole.
\item A PTA identifies a potential supermassive black hole binary by its gravitational-wave signature.  Although the initial localization is crude, coordinated multi-wavelength searches guided by the tentatively measured orbital frequency can locate and confirm the host galaxy and enable tests of galaxy properties and accretion disk astrophysics.
\item A long-awaited supernova in our Galaxy or the Local Group can be detected by next-generation MeV neutrino detectors and potentially by the GW detector network, depending on the supernova type.  Either or both would complement optical observations of the supernova light curve and yield information about core-collapse astrophysics, nuclear physics, and shock acceleration.
\end{itemize}
For a richer narrative of each of these examples, describing the interplay of coordinated observations, analysis and modeling that can be enabled by advanced cyberinfrastructure, see the Appendix: Visions of MMA Discovery.  While the facilities supplying the observational data are already operating or else planned, each scenario uses many coordination and rapid analysis capabilities which do not currently exist, but which could be developed with the help of new investment in cyberinfrastructure for MMA.

\section{Cyberinfrastructure Challenges for MMA}
\label{s:challenges}
By its nature, MMA requires different scientific teams to bring together their observational resources, data, analysis and modeling tools, and expertise with different instruments, different data formats, different analysis protocols, and different cultures.  Given the highly heterogeneous data as well as differing collaboration styles, working together effectively requires a mutual framework for collaboration, involving communication among individual humans, projects, and machines. 

These groups require the ability to form flexible collaborations on short time scales, to share data, codes and other digital objects in real time, and to self-organize into spontaneous collaborations of varying scope, both in terms of the instruments and people involved and the digital objects they share. As well as digital objects, different groups have existing hardware infrastructures and computing resources which may need to be shared.

On the organizational level, teams may need to share proprietary data and information governed by memoranda of understanding (MoUs). This may be facilitated by providing a means to hold proprietary data in ``escrow''.   On the technical level, MMA will benefit from intelligent scheduling and coordination of follow-up observations, as well as on-demand modeling to interpret the early data and inform further observations and analysis. This will be especially important when there are a multitude of astrophysical source candidates to be confirmed and fully characterized.

The organizational requirements will need to be met by technical implementations and controls. MoUs may define the scope of collaboration between projects, but practical implementation requires that there be flexible and dynamic ways of defining collaboration members, access controls on escrowed data, and means to enforce that pre-registered analyses are not altered. Tools that allow for scheduling of observations and sharing of relevant information, both in human-readable and machine-readable form, will be required. Storage models with sortable and filterable interfaces for candidates will allow easier prioritization and comprehension of large event candidate pools.

In MMA we are considering (by definition!) sources that must be detected across multiple qualitatively different facilities, in some cases simultaneously or with low latency. In general these facilities will have different uptimes, different sensitivities to weather and geography, and different noise properties, angular resolutions, and sources of confusion.  Therefore every MMA source of interest will be observed with a different footprint in time, cadence, wavelength bandpass, and precision as it is followed up by heterogeneous facilities. And yet among our most important questions about these sources are statistical questions of the form: What is the true, underlying population of these sources? What are the distributions in fluence, timescale, and energy partitions (to different messengers) across domains?

There are no simple solutions to these problems but provides significant opportunities for collaborative research between MMA and a broad range of computational disciplines. The success of the MMA cyberinfrastructure will depend on both theoretical advances in how we represent data and the information they contain as well as the development of applications and systems that can scale to the complexity and size of data from a heterogenous network of astrophysical facilities.  Close interaction between MMA and programs within the NSF such as the Office of Advanced Cyberinfrastructure (OAC), Cyber-Human Systems  (CHS),  and Information \& Intelligent Systems (IIS) can provide the necessary collaboration among data science, statistics, and astrophysics.  Desired cyberinfrastructure capabilities include:
\begin{itemize}
\item Pre-negotiated data exchange service between facilities and observing resource allocation to enable rapid analysis and follow-up 
\item A framework to facilitate joint analysis, and to enable various teams to work together, while respecting their different scientific cultures (e.g., data access and rights, publication policy, etc.)
\item Real-time decision-making in event observing and follow-up
\item Coordination of observing resources through exchanges, marshals, or Telescope Observation Managers (TOMs)
\item Capability-based access controls, authentication systems, dynamically-deployed and managed storage services integrated with resource schedulers to ensure that data on shared resources matches the requirements of the data analyses which rely on it
\item Sustainable, long-term archival storage of data capable of dynamical updates and heterogeneous data structures
\item Collaboration-enabling services (wikis, mailing lists, scheduling, calendaring, code repositories, data stores) and management platforms
\item Standardization and central management of data sets
\item Means to hold proprietary data in ``escrow'', or by a protocol for pre-registration of analyses to be executed on data as it becomes available
\item Machine-readable standards and protocols
\item Communication software:  human-to-human, machine-to-machine, and human-to-machine
\item Frameworks for likelihood optimization and Bayesian inference
\item Scalable compute systems (servers, compute clusters) for analyzing and serving data (centralized or on the cloud)
\end{itemize}

\section{Opportunities for Computer Science and Data Science}
\label{s:compsci}
There is an exciting cross-disciplinary whitespace---in the interface between existing algorithms, data, and novel questions---that cannot be addressed by existing hardware, software, and analysis approaches. Indeed, we believe that harnessing the data revolution for MMA demands deep collaborations with researchers from a diverse set of disciplines including computer scientists, data scientists, and statisticians to lead to the invention of novel approaches to advance MMA scenarios, and to provide the inspiration for new, futuristic MMA scenarios inspired by new computing and data approaches.

Areas of data science opportunities that were identified at the workshop include:
\begin{itemize}
\item Innovations in time-series analysis
\item Machine learning and deep learning innovations
\item Purpose-built hardware for real-time inferencing
\item Allowing for inferencing and compute on censored data and differential privacy
\item Uncertainty quantification and predictive modelling 
\item Tools for dealing with missing data and imbalanced datasets
\item Cyberinfrastructure and workflow scheduling technologies to manage and coordinate heterogeneous computing models, resources, and data from diverse instruments and projects
\end{itemize}

\section{Broader Impacts Opportunities}
\label{s:broader}
\subsection{Diversity}

This is a once-in-a-generation opportunity to build a new field. Since MMA sits at the interface between data science, astronomy, and engineering,  we need to draw upon an exceptionally broad and global talent pool. Creating this vibrant and inclusive new field requires participation from diverse specialties, from groups traditionally underrepresented in STEM, and from a wide range of cultural backgrounds.  By embracing a commitment to broadening participation from the start, we will set a tone that maximizes the potential for transformative science.  Potential activities include:
\begin{itemize}
\item Hosting regular ``matchmaking'' workshops for MMA researchers and groups of undergraduate science majors seeking REUs or other summer research opportunities.
\item Hosting Bridge programs available for underrepresented groups (including, but not limited to gender, race, disability, and socioeconomic status) that grant a MA degree in order to build a strong academic foundation and research skills that foster a successful transition to a PhD program.
\item Collaboration opportunities for faculty from underrepresented groups through partnerships with scientists at MMA facilities and institutions.
\end{itemize}

\subsection{Public Outreach}

While each observatory or project will reach out to the public to share its own angle on astronomy, physics or computer-assisted discovery, the multidisciplinary nature of MMA naturally creates opportunities for "big picture" outreach, highlighting what can be achieved with smart collaboration among people with diverse skills.  Cyberinfrastructure developed for the core scientific mission also can support citizen science platforms to reach a large audience using the appeal of contributing to ``big-data'' science.

\section{Scientific Community Needs}
\label{s:community}
\subsection{Expertise Sharing and Curation}
\label{ss:expertise}

Multi-messenger astrophysics is an intrinsically interdisciplinary endeavor.  It provides immense opportunity for a diverse community of scientists to collaborate and advance knowledge in a way that is bigger than the sum of the parts.  For MMA to be successful, there needs to be a path to facilitate the sharing of expertise and potentially curate expertise that is critical for successful multi-messenger projects.  Such expertise falls roughly into three categories: 

{\bf Domain-specific expertise}: The community needs a hub for the exchange of domain-specific expertise through hosting topical workshops and by serving as a gateway to domain-specific resources that may be well known within individual communities, while not being broadly recognized beyond these communities.

{\bf Multi-messenger data analysis}: The community needs a core in statistical inference methods, data science and cyberinfrastructure, including experts on call for researchers hoping to advance their own expertise in these areas that directly impact multi-messenger work.   
The community also needs a strategy to prioritize critical computer simulation development (general relativistic magnetohydrodynamics, radiation and neutrino transport, chemical processes etc.) and provide support and training.

{\bf Software engineering and cyberinfrastructure}: The community needs a core set of research software engineers who are broadly recognized as epitomizing software engineering best practices and are adept at conveying these best practices to others. Although there have been significant advances in software engineering practices (typically in the context of large survey programs such as the LSST or WFIRST),
data analysis and infrastructure software in scientific domains tends to be of inconsistent quality since it is often written by scientists with little or no software engineering skills.  Often this software is critically relied upon but there is no path for sustainability and curation of the software or even the function that it provides. Shared, high quality expertise in software engineering has the power to greatly improve the quality of domain science software.

\subsection{Data Stewardship and Sustainability}
\label{ss:stewardship}

Individual MMA projects generally take responsibility for their own live archives and databases, but to date no one is responsible for the MMA archives for completed projects, for their documentation and software, or for the curation of software, services, and data sets generated by the broader MMA community, including correlative and follow-up observations. Further, there does not yet exist a central clearinghouse of normalized data across MMA for researchers to peruse, explore, and develop against. Such an entity could provide enormous value to the MMA community by bringing together disparate data sets, maintaining them in perpetuity, building a range of interfaces, and thus lowering the bar to entry for researchers.
	
As existing MMA projects reach completion or wish to make public data releases, there are few options for making these potentially large datasets publicly available. A single archive, modeled after data archives like the Mikulski Archive for Space Telescopes (MAST) or the Infrared Processing and Analysis Center (IPAC), could capture project expertise in data products, software, and documentation during its primary operations phase. This would keep data accessible and scientifically usable by non-experts during and beyond the lifetime of an individual project. Once available in a larger catalog of MMA data products, these data would enable MMA researchers to make new discoveries, either alone or in combination with the data from other projects. 

However, this also presents challenges.  We need to build expertise across an extremely wide and evolving landscape, with heterogeneous, multi-wavelength/multi-messenger datasets that require developing brand-new algorithms, analysis, and models.  In addition to the technical and scientific expertise, we need to pass on best practices  and ensure a cooperative and inclusive community.  Much of the current knowledge in this field is passed down in limited, informal training that cannot scale up to large groups of new students, especially given the diverse global nature of next-generation MMA.  Although webinars and workshops are helpful, they are not enough to build a new field.  We have the opportunity to establish international multi-disciplinary cohorts that can learn from each other and help spread expertise more quickly.

\section{Conclusions, and the Need for an Institute}
\label{s:center}
Ultimately, improving the speed and efficiency of obtaining, analyzing and characterizing observational data will lead to better science and open up further opportunities. Multi-messenger astrophysics therefore presents both unique challenges \emph{and} opportunities spanning the disciplines of physics, astronomy and astrophysics. Further, its big-data character requires us to leverage significant cyberinfrastructure expertise, tools and techniques, as well as developing new algorithms and techniques where needed.

The workshop attendees judged that an Institute for Multi-Messenger Astrophysics would be ideally suited to address the long-term challenges in this emerging field.  A sustained effort is needed to focus on expertise and software infrastructure connections, as well as to identify opportunities for new applications of computer science.  We envision a distributed institute for MMA advancement that brings together researchers with different expertise in order to provide cyberinfrastructure that will enable and support critical aspects needed in this era of MMA.  This includes:
\begin{itemize}
\item A centralized, community-accepted platform for machine-learning-enabled event handling and facility coordination
\item A long-term storage facility for the management of MMA data sets
\item A knowledge base for statistical techniques that will standardize MMA analysis across the various messengers, and uphold a strong standard of analysis and results
\item Provide and support communication tools and identity management to serve a heterogeneous community of researchers
\item Provide a point of contact for distributed outreach and collaboration opportunities
\item Availability of workshops to further machine learning and deep learning research in this field
\item Availability of fellows who can bridge the many disciplines required for MMA analysis (software, distributed computing, statistics, management, science, etc.)
\item Software engineering and cyberinfrastructure expertise could be provided through direct consultation, regular workshops, or as a gateway to knowledge for those who wish to learn independently
\end{itemize}
No single institution has the scope of expertise to provide, develop and support all of the capabilities that will be needed to carry out an optimal MMA program.  Indeed, the MMA effort draws its strength from coordinating the activities of different groups to make something greater than the sum of the parts.  The institute will support existing and near-future facilities and projects, that have been funded by the National Science Foundation and other agencies, to enable additional science.  It will enable efficient transfer of expertise and training opportunities for many scientists in different fields.

With this summary, we have aimed to lay out the topics and ideas that were discussed in the workshop, and to establish context for further community engagement.  A natural next step toward Harnessing the Data Revolution for Multi-Messenger Astrophysics would be to conduct a design study that engages the community to work out a vision for the future in more detail.

\section{Acknowledgements}
The Cyberinfrastructure for Multi-Messenger Astrophysics workshop was supported by National Science Foundation grant PHY-1838082.
Workshop participants also acknowledge support from NSF grants PHY-1412633 (DF), PHY-1505230 (IT), and others.
SBS is a member of the NANOGrav project, which is supported by NSF Physics Frontier Center award number 1430284.

\clearpage
\setcounter{secnumdepth}{0}
\section{Appendix: Visions of MMA Discovery}
\label{s:visions}
We present here a few examples of the vision we have for MMA in the 2020s, as a way of motivating the potential challenges and opportunities that will arise.

\subsection{Binary neutron star inspiral detected prior to merger}

Early in the 2020s the LIGO and VIRGO detectors will be operating at design sensitivity, in coordination with science operations of KAGRA, and the joint network will be capable of $\sim$100s pre-merger detection of binary neutron star (BNS) coalescence events out to a distance of 200 Mpc. We therefore envision the following sequence of events: 

LIGO+VIRGO+KAGRA (LVK) identify a 80 Mpc-distant BNS at 100s pre-merger, leading to an “early warning” pre-merger alert; 
A GW transient alert and $\sim$50 deg$^{2}$ localization is automatically generated by the LVK data analysis system and distributed globally to a broad range of interested observers and theorists; 
Available telescopes with approved programs receive the alert and negotiate via agent-based methods to observe the localization in “divide and conquer” fashion, imaging in multiple optical bands; 
Simultaneously, the on-orbit SVOM spacecraft re-points to observe the localization with its high-energy instruments; 
Within one second of the BNS merger, a bright hard X-ray transient is detected by SVOM and reported as a GW counterpart, via a revision to the original LVK alert;
Just one second later, LVK updates the alert with a refined localization from the full GW signal, and notes that it is consistent with the SVOM burst; 
Seconds later a further update from a robotic telescope (one of those that agreed to observe the GW localization) reports the presence of a V~12 mag optical counterpart; 
Available spectroscopic facilities with approved programs note the bright BNS optical counterpart and negotiate to observe in the optical and near-infrared, again via distinct facilities; 
Over subsequent hours, the Earth rotates and telescope programmatic priorities shift, leading to hand-over of spectroscopic observations to additional telescopes without any gap in coverage; 
The rise of the red kilonova component is reported by a robotic imaging facility and triggers an approved space telescope program to observe the red kilonova over 1 $\mu$m to 5 $\mu$m; 
An observing agent configured by a BNS theory/modeling group notes the lack of sensitive mm-wavelength observations of the rising kilonova and broadcasts a call for observations;
Subsequent ALMA+ observations register the first early detection of the kilonova and provide unique constraints on the BNS outflow;  
Follow-up modelling on a timescale of days yields predictions for the infrared evolution of the kilonova that is used to design the subsequent NIR/IR spectroscopic campaign so as to test models of the nuclear equation of state.

\subsection{High-energy neutrinos from a tidal disruption event}

Joint operations of a full-strength or expanded KM3NET with IceCube-Gen2 would provide unique sensitivity to TeV--PeV neutrino transient populations. We envision the following scenario: 

Through MMA cyberinfrastructure, a neutrino coincidence between KM3NET and IceCube-Gen2 is identified, which exhibits an estimated false alarm rate of 3 yr$^{-1}$; 
a joint localization of 0.4 deg$^{2}$ is derived, and the neutrino coincidence alert is distributed to a broad group of interested observers; 
available telescopes with approved programs negotiate to perform a coordinated imaging search for optical counterparts; 
an agent configured by a group of IceCube members annotates the alert and redistributes it with a comment that the alert localization includes an Abell galaxy cluster and several of its component galaxies, including the central type-cD galaxy; 
the possible association of the neutrino with a galaxy cluster triggers an approved program for deep imaging and X-ray observations using the eROSITA mission; 
as the Earth rotates, optical imaging observations are continued at robotic facilities around the globe; 
after 12 hours, a new X-ray source, superposed on the nuclear regions of a dwarf galaxy in the cluster, is identified in eROSITA data, and the X-ray source and its likely host galaxy are reported as an alert update; 
automated MMA systems annotate the alert with an optical history of the galaxy from LSST observations, showing ~0.1 mag stochastic variability; 
the agent for an approved program to observe nuclear high-energy neutrino transients notes that the updated position is consistent with the nucleus of one of the member galaxies of the cluster, and requests deep NIR IFU spectroscopy of the galaxy's nuclear region; 
almost simultaneously, robotic facilities update the alert to note that the galaxy nucleus has brightened by $\sim$1 mag since the first observational epoch; 
together X-ray, optical imaging, and NIR spectroscopic data establish the existence of a tidal disruption event (TDE) as the high-confidence source of the neutrino coincidence, leading to novel constraints on hadronic acceleration processes in TDEs. 

\subsection{Supermassive black hole binary identification}

Joint MMA
identification of an inspiralling supermassive black hole binary
(SMBHB) would allow a unique understanding of SMBHB evolution and the
connection to the larger galactic host environment. PTA observations
are likely to identify a number of continuous-wave sources
representing the early inspiral phase of SMBHBs. They will provide a
crude localization (up to a few thousand deg$^2$), and will provide a
direct measurement of frequency, however with degeneracies in mass,
distance, and eccentricity. Identifying the host galaxy of such a
system can break the mass/distance degeneracy through a redshift
measurement, and thus allow a precision mass measurement for the
binary that can be directly compared to host galaxy properties. Once a
continuous wave is detected by PTAs, we can statistically limit the
target galaxies based on a given mass/distance ratio to create a host
candidate catalog; Multi-wavelength searches with optical, X-ray,
radio, and other facilities can work together to identify SMBHB
markers in those galaxies (e.g. periodic variability, disturbed
galactic systems, AGN); deep searches with VLBI and spectral lines may
identify kinematic or visual binaries; when a host is identified,
$M_{\rm BH}-$host relations can be measured, circumbinary disk
evolution can be studied, and the mechanics of SMBHB inspiral will be
dissected. PTAs may also see the merger event through the
gravitational-wave memory effect, which may likewise result in a
transient. It is worth noting that the time scale involved in MMA for
the SMBHBs described here is slower, as the systems can endure from
months to decades, depending on the stage of inspiral at which they
are detected.

\subsection{Galactic supernova}

The global network of MeV neutrino observatories, coordinated by SNEWS, has kept watch for the next Galactic supernova for 20 years. Completion of the next generation of detectors, in particular Hyper-Kamiokande and Watchman, will extend this sensitivity out to M31, more than doubling the expected rate of core-collapse supernova detections. With advanced GW facilities now in operation, a Galactic supernova in particular raises real prospects of a triply-detected event, with a GW trigger and MeV neutrinos preceding the initial “shock breakout” EM signature by minutes (type Ibc) to hours (type II). Subsequent observations and interpretation of the GW and neutrino signals would yield insights into the physics of core collapse, nuclear physics, fundamental physics, and the astrophysics of shocks and shock acceleration.

\end{document}